\documentclass[12pt,letterpaper]{article}%
\usepackage{setspace}
\usepackage{amssymb}
\usepackage{amsmath}
\usepackage{amsfonts}
\usepackage{graphicx}
\usepackage{color}%
\ifx\pdfoutput\relax\let\pdfoutput=\undefined\fi
\newcount\msipdfoutput
\ifx\pdfoutput\undefined\else
\ifcase\pdfoutput\else
\ifx\paperwidth\undefined\else
\ifdim\paperheight=0pt\relax\else\pdfpageheight\paperheight\fi
\ifdim\paperwidth=0pt\relax\else\pdfpagewidth\paperwidth\fi
\fi\fi\fi
\begin{document}

\title{Understanding the Hastings Algorithm }
\author{David D. L. Minh\\Illinois Institute of Technology\\BCHS Chemistry Division\\3101 South Dearborn St.\\Chicago, IL 60616
\and Do Le (Paul) Minh\\Department of ISDS, California State University, Fullerton\\CA\ 92831, USA\\dminh@fullerton.edu}
\date{}
\maketitle

\section*{Abstract}

The Hastings algorithm is a key tool in computational science. While
mathematically justified by detailed balance, it can be conceptually difficult
to grasp. Here, we present two complementary and intuitive ways to derive and
understand the algorithm. In our framework, it is straightforward to see that
the celebrated Metropolis-Hastings algorithm has the highest acceptance
probability of all Hastings algorithms.

\bigskip

Keywords: Hastings algorithm; Metropolis-Hastings algorithm; Markov chain
Monte Carlo; Simulation.

Mathematics Subject Classification: Primary 65C05; Secondary 78M31,
80M31.\bigskip

\section{Introduction}

\subsection{The Hastings algorithm (HA)}

The Hastings algorithm (HA) (Hastings, 1970) is a stochastic sampling
technique widely used throughout computational science. As a Markov Chain
Monte Carlo method, HA does not attempt to generate a sequence of independent
samples from a \textquotedblleft target distribution\textquotedblright%
\ $\pi(\cdot)$, defined on the state space $(E,\mathcal{E})$, but rather a
Markov chain $\left\{  X_{n},n=1,2,3,...\right\}  $ having $\pi(\cdot)$ as its
invariant distribution. Although variates in the chain are not independent,
they may nonetheless be used to estimate statistical expectations with respect
to $\pi(\cdot)$. (In a slight abuse of notation, we will often use the same
symbol to denote both a measure and its density function.)

In many applications, the target distribution takes the form $\pi\left(
\cdot\right)  =p\left(  \cdot\right)  /P$, where the normalizing constant
$P=\int_{E}p\left(  x\right)  dx$ is unknown. We call $p(\cdot)$ the
un-normalized target distribution and $\pi(\cdot)$ the normalized one. If $x$
is a variate generated from $\pi(\cdot)$, we may interchangeably write
$x\sim\pi(\cdot)$ or $x\sim p(\cdot)$.

Let $U(0,1)$ represent the uniform distribution on $(0,1)$. In order to use
all subsequently described algorithms, given $X_{n}=x$, we require a
\textquotedblleft proposal density\textquotedblright\ $\gamma(\cdot|x)$ which
may (or may not) depend on $x$, and whose variates can be generated by other means.

Given $X_{n}=x\sim\pi(\cdot)$, we can generate $X_{n+1}\sim\pi(\cdot)$
by\bigskip

\textbf{Algorithm }$\mathbf{HA}$ (Hastings)

\begin{enumerate}
\item[HA1.] generate $y\sim\gamma(\cdot|x)$ and $r\sim U(0,1)$

\item[HA2.] if $r\leq\alpha_{HA}(x,y)$, output $X_{n+1}=y$

\item[HA3.] else, output $X_{n+1}=x$\bigskip
\end{enumerate}

\noindent where $\alpha_{HA}(x,y)$ is the Hastings' \textquotedblleft
acceptance probability,\textquotedblright\thinspace defined in terms of a
symmetric function $s(\cdot,\cdot)$ that satisfies the following condition:
For all $x,y\in E$,
\begin{equation}
0\leq\alpha_{HA}(x,y)=s(x,y)\left(  1+\frac{p(x)}{\gamma(x|y)}\frac
{\gamma(y|x)}{p(y)}\right)  ^{-1}\leq1. \label{01}%
\end{equation}
(In Equation (6) in Hastings, 1970, this condition was expressed in terms of
the normalized $\pi(\cdot)$, rather than the un-normalized $p(\cdot)$.)

\subsection{Some special forms of the Hastings algorithm}

\subsubsection{The Metropolis-Hastings algorithm (MH)}

HA was introduced as a generalization of the previously known Metropolis
(1953) and Barker (1965) algorithms. In the celebrated paper by Metropolis,
Rosenbluth, Rosenbluth, Teller and Teller (1953), the proposal densities are
assumed to be symmetric (that is, $\gamma(x|y)=\gamma(y|x)$) and the
acceptance probability in Step HA2 is,%
\[
\alpha_{MT}(x,y)=\min\left\{  \frac{p(y)}{p(x)},1\right\}  .
\]
Hastings generalized the Metropolis algorithm into the well-known
Metropolis-Hastings algorithm (MH) by setting $s(x,y)=s_{MH}(x,y)$, where,%
\begin{align}
s_{MH}(x,y)  &  =\left\{
\begin{array}
[c]{ll}%
1+\frac{p(x)}{\gamma(x|y)}\frac{\gamma(y|x)}{p(y)} & \text{if }\frac
{\gamma(x|y)}{p(x)}\frac{p(y)}{\gamma(y|x)}\geq1\text{ }\\
1+\frac{p(y)}{\gamma(y|x)}\frac{\gamma(x|y)}{p(x)} & \text{if }\frac
{\gamma(y|x)}{p(y)}\frac{p(x)}{\gamma(x|y)}\geq1
\end{array}
\right. \nonumber\\
&  =\left(  1+\frac{p(x)}{\gamma(x|y)}\frac{\gamma(y|x)}{p(y)}\right)
\min\left\{  \frac{\gamma(x|y)}{p(x)}\frac{p(y)}{\gamma(y|x)},1\right\}  .
\label{02}%
\end{align}
The acceptance probability $\alpha_{HA}(x,y)$ in Equation (\ref{01}) then
becomes the well-known MH acceptance probability (Chib and Greenberg, 1995 and
Tierney, 1994):
\begin{equation}
\alpha_{MH}(x,y)=\min\left\{  \frac{\gamma(x|y)}{p(x)}\frac{p(y)}{\gamma
(y|x)},1\right\}  . \label{03}%
\end{equation}

\subsubsection{The Barker algorithm (BK)}

Barker (1965) proposed the following acceptance probability, which uses the
symmetric proposal densities $\gamma(x|y)=\gamma(y|x)$,%
\[
\alpha_{BK}^{(s)}(x,y)=\left(  1+\frac{p(x)}{p(y)}\right)  ^{-1},
\]
which Hastings generalized by setting $s(x,y)=1$ in Equation (\ref{01}):%
\begin{equation}
\alpha_{BK}(x,y)=\left(  1+\frac{p(x)}{\gamma(x|y)}\frac{\gamma(y|x)}%
{p(y)}\right)  ^{-1}. \label{04}%
\end{equation}
We will subsequently refer to the Hastings algorithm with the acceptance
probability $\alpha_{BK}$ as the Barker algorithm (BK).

\subsubsection{Another special form of HA}

As another example of HA, consider the case where $s(x,y)$ takes the following
symmetric form:%
\begin{align}
s(x,y)  &  =\min\left(  \frac{\gamma(x|y)}{p(x)},1\right)  \min\left(
\frac{\gamma(y|x)}{p(y)},1\right)  \left(  \frac{p(x)}{\gamma(x|y)}%
+\frac{p(y)}{\gamma(y|x)}\right) \label{4c}\\
&  =\min\left(  \frac{\gamma(x|y)}{p(x)},1\right)  \min\left(  \frac
{p(y)}{\gamma(y|x)},1\right)  \left(  1+\frac{p(x)}{\gamma(x|y)}\frac
{\gamma(y|x)}{p(y)}\right)  .\nonumber
\end{align}
Substituting this form of $s(x,y)$ into Equation (\ref{01}) results in the
following acceptance probability for all $x,y\in E$:%
\begin{equation}
\min\left(  \frac{\gamma(x|y)}{p(x)},1\right)  \min\left(  \frac{p(y)}%
{\gamma(y|x)},1\right)  \leq1. \label{4b}%
\end{equation}

\subsection{The detailed balance}

To prove that a Markov chain $\left\{  X_{n},n=1,2,3,...\right\}  $ has the
invariant distribution $\pi(\cdot)$, it is sufficient to show that its
transition kernel $P(\cdot|\cdot)$ satisfies detailed balance (which is also
called the \textquotedblleft reversibility condition\textquotedblright) with
respect to $p(\cdot)=P\pi\left(  \cdot\right)  $; that is, for all $x,y\in E,$%
\[
p(x)P(y|x)=P(x|y)p(y).
\]
In this paper, all transition kernels can be expressed in two parts,
\[
P(y|x)=r_{1}(y|x)+I(x=y)r_{2}(y|x),
\]
where $I(a)=1$ if $a$ is true, $0$ otherwise. Because $p(x)I(x=y)r_{2}(y|x)=$
$p(y)I(x=y)r_{2}(x|y)$, for notational simplicity, we only prove detailed
balance for $x\neq y$, omitting the second part.

For HA, the transition kernel for all $x,y\in E$ is,
\begin{equation}
P_{HA}(y|x)=\alpha_{HA}(x,y)\gamma(y|x)+I(x=y)\left[  \int_{E}(1-\alpha
_{HA}(x,z))\gamma(z|x)dz\right]  . \label{05}%
\end{equation}
The first term is the probability of the proposed variate $y\sim\gamma
(\cdot|x)$ being accepted (the chain moves to $y$). The term inside the
integration is probability of the proposed variate $z\sim\gamma(\cdot|x)$
being the rejected (the chain remains at $x$). Thus $P_{HA}(y|x)$ satisfies
detailed balance with respect to $p(\cdot)$ because, from Equation (\ref{01}),
for all $x,y\in E$ and $x\neq y$,%
\begin{align*}
p(x)P_{HA}(y|x)  &  =p(x)\alpha_{HA}(x,y)\gamma(y|x)\\
&  =p(x)s(x,y)\frac{p(y)\gamma(x|y)}{p(x)\gamma(y|x)+p(y)\gamma(x|y)}%
\gamma(y|x)\\
&  =P_{HA}(x|y)p(y).
\end{align*}

While verifying that HA satisfies detailed balance is simple, conceptually
understanding it is much harder. In a paper interpreting MH geometrically,
Billera \& Diaconis (2001) wrote, \textquotedblleft The algorithm is widely
used for simulations in physics, chemistry, biology and statistics. It appears
as the first entry of a recent list of great algorithms of 20th-century
scientific computing [4]. Yet for many people (including the present authors)
the Metropolis-Hastings algorithm seems like a magic trick. It is hard to see
where it comes from or why it works.\textquotedblright\ (Reference [4] refers
to Dongarra and Sullivan, 2000.) If it is hard to conceptually understand the
development of MH, it is even harder to visualize the more general HA.

In this paper, we provide two complementary and intuitive derivations of the
Hastings algorithm. First, we present a new form of the acceptance probability
in the next section.

\section{Algorithm $M$}

\subsection{Algorithm $M$}

Given $X_{n}=x\sim\pi(\cdot)$, $X_{n+1}\sim\pi(\cdot)$ can be generated
by\bigskip

\textbf{Algorithm }$\mathbf{M}$

\begin{enumerate}
\item[M1.] generate $y\sim\gamma( \cdot|x) $ and $r\sim U( 0,1) $

\item[M2.] if $r\leq\alpha_{M}(x,y)$, output $X_{n+1}=y$

\item[M3.] else, output $X_{n+1}=x$\bigskip
\end{enumerate}

\noindent in which the acceptance probability $\alpha_{M}(x,y)$ is, for all
$x$, $y\in E$,%
\begin{equation}
\alpha_{M}(x,y)=\min\left\{  \frac{k(x,y)\gamma(x|y)}{p(x)},1\right\}
\min\left\{  \frac{p(y)}{k(x,y)\gamma(y|x)},1\right\}  \leq1, \label{06}%
\end{equation}
where$\ k\left(  \cdot,\cdot\right)  :E\times E\rightarrow R>0$ is a
\emph{symmetric} function.

Similar to Equation (\ref{05}), the transition kernel of Algorithm $M$ is, for
all $x,y\in E$,
\[
P_{M}(y|x)=\alpha_{M}(x,y)\gamma(y|x)+I(x=y)\left[  \int_{E}(1-\alpha
_{M}(x,z))\gamma(z|x)dz\right]  .
\]
$P_{M}(y|x)$ satisfies detailed balance with respect to $p(\cdot)$: For all
$x,y\in E$ and $x\neq y$,
\begin{align*}
&  p(x)P_{M}(y|x)=p(x)\alpha_{M}(x,y)\gamma(y|x)\\
&  =p(x)\min\left\{  \frac{k(x,y)\gamma(x|y)}{p(x)},1\right\}  \min\left\{
\frac{p(y)}{k(x,y)\gamma(y|x)},1\right\}  \gamma(y|x)\\
&  =\min\left\{  k(x,y)\gamma(x|y),p(x)\right\}  \min\left\{
p(y),k(x,y)\gamma(y|x)\right\}  \frac{1}{k(x,y)}\\
&  =P_{M}(x|y)p(y).
\end{align*}

If $k(x,y)$ is a positive constant $k$, then $p(\cdot)/k$ is just another
un-normalized distribution corresponding to $\pi(\cdot)$ and the acceptance
probability $\alpha_{M}(x,y)$ in Equation (\ref{06}) is the same as that in
Equation (\ref{4b}). So, for the rest of this paper, we exclude the case in
which $k(x,y)=k$.

For all $x$, $y\in E$, we define,
\[
L(x,y)=\min\left\{  \frac{p(x)}{\gamma(x|y)},\frac{p(y)}{\gamma(y|x)}\right\}
\text{ and }H(x,y)=\max\left\{  \frac{p(x)}{\gamma(x|y)},\frac{p(y)}%
{\gamma(y|x)}\right\}  .
\]
When $L(x,y)<k(x,y)<H(x,y)$, from Equation (\ref{06}),
\begin{align}
\alpha_{M}(x,y)  &  =\left\{
\begin{array}
[c]{ll}%
1 & \text{if }\frac{p(x)}{\gamma(x|y)}<k(x,y)<\frac{p(y)}{\gamma(y|x)}\\
\frac{\gamma(x|y)}{p(x)}\frac{p(y)}{\gamma(y|x)} & \text{if }\frac
{p(y)}{\gamma(y|x)}<k(x,y)<\frac{p(x)}{\gamma(x|y)}%
\end{array}
\right. \nonumber\\
&  =\min\left\{  \frac{\gamma(x|y)}{p(x)}\frac{p(y)}{\gamma(y|x)},1\right\}  .
\label{07b}%
\end{align}
Thus the acceptance probability $\alpha_{M}(x,y)$ may be expressed as a
piecewise function that depends on the relationship between $k(x,y)$,
$L(x,y)$, and $H(x,y)$:
\begin{equation}
\alpha_{M}(x,y)=\left\{
\begin{array}
[c]{ll}%
\frac{p(y)}{k(x,y)\gamma(y|x)} & \text{if }k(x,y)\geq H(x,y)\\
\min\left\{  \frac{\gamma(x|y)}{p(x)}\frac{p(y)}{\gamma(y|x)},1\right\}  &
\text{if }L(x,y)<k(x,y)<H(x,y)\\
\frac{k(x,y)\gamma(x|y)}{p(x)} & \text{if }k(x,y)\leq L(x,y)
\end{array}
\right.  \label{11}%
\end{equation}

From Equations (\ref{03}) and (\ref{07b}), it is clear that MH is a special
case of Algorithm $M$ when $L(x,y)<k(x,y)<H(x,y)$.

BK, with acceptance probability $\alpha_{BK}(x,y)$ in Equation (\ref{04}), can
also be shown to be a special case of Algorithm $M$: We set,%
\[
k(x,y)=\frac{p(x)}{\gamma(x|y)}+\frac{p(y)}{\gamma(y|x)}\geq\max\left\{
\frac{p(x)}{\gamma(x|y)},\frac{p(y)}{\gamma(y|x)}\right\}  =H(x,y).
\]
Then from Equation (\ref{11}),%
\[
\alpha_{M}(x,y)=\frac{p(y)}{\gamma(y|x)}\left(  \frac{p(x)}{\gamma(x|y)}%
+\frac{p(y)}{\gamma(y|x)}\right)  ^{-1}=\left(  1+\frac{p(x)}{\gamma
(x|y)}\frac{\gamma(y|x)}{p(y)}\right)  ^{-1}=\alpha_{BK}(x,y).
\]

We can also set,
\begin{align*}
k(x,y)  &  =\left(  \frac{\gamma(x|y)}{p(x)}+\frac{\gamma(y|x)}{p(y)}\right)
^{-1}\\
&  \leq\left(  \max\left\{  \frac{\gamma(x|y)}{p(x)},\frac{\gamma(y|x)}%
{p(y)}\right\}  \right)  ^{-1}=\min\left\{  \frac{p(x)}{\gamma(x|y)}%
,\frac{p(y)}{\gamma(y|x)}\right\}  =L(x,y),
\end{align*}
to obtain the same BK acceptance probability from Equation (\ref{11}):
\[
\alpha_{M}(x,y)=\frac{\gamma(x|y)}{p(x)}\left(  \frac{\gamma(x|y)}{p(x)}%
+\frac{\gamma(y|x)}{p(y)}\right)  ^{-1}=\left(  1+\frac{p(x)}{\gamma
(x|y)}\frac{\gamma(y|x)}{p(y)}\right)  ^{-1}=\alpha_{BK}(x,y).
\]

\subsection{Algorithm $M$ and HA}

We now show that HA and Algorithm $M$ are equivalent. First, we show that the
former is a special case of the latter. We then show that the latter is a
special case of the former.

\subsubsection{HA is a special case of Algorithm $M$}

HA is a special case of Algorithm $M$ if, for any acceptance probability
$\alpha_{HA}(\cdot,\cdot)$ in HA, expressed in terms of $s(\cdot,\cdot)$, we
can find the same acceptance probability $\alpha_{M}(\cdot,\cdot)$ in
Algorithm $M$: For each $s(\cdot,\cdot)$ satisfying the Hastings condition
(\ref{01}), we define the following symmetric function,%
\begin{equation}
M_{s}(x,y)=\frac{1}{s(x,y)}\left(  \frac{p(x)}{\gamma(x|y)}+\frac{p(y)}%
{\gamma(y|x)}\right)  =\frac{1}{s(x,y)}\left(  1+\frac{p(x)}{\gamma(x|y)}%
\frac{\gamma(y|x)}{p(y)}\right)  \frac{p(y)}{\gamma(y|x)}\geq\frac
{p(y)}{\gamma(y|x)}. \label{13}%
\end{equation}
Because $M_{s}(x,y)$ is symmetric, we also have $M_{s}(x,y)\geq p(x)/\gamma
(x|y)$; hence $M_{s}(x,y)\geq H(x,y)$. Now letting $k(x,y)=M_{s}(x,y)$ in
Equation (\ref{11}), we find in Algorithm $M$ the same acceptance probability
as that defined by $s(x,y)$ in HA:
\begin{equation}
\alpha_{M}(x,y)=\frac{p(y)}{M_{s}(x,y)\gamma(y|x)}=s(x,y)\left(  1+\frac
{p(x)}{\gamma(x|y)}\frac{\gamma(y|x)}{p(y)}\right)  ^{-1}=\alpha_{HA}(x,y).
\label{14}%
\end{equation}

For example, if $s(x,y)$ takes the form of Equation (\ref{4c}), then Equation
(\ref{13}) yields%
\begin{align*}
M_{s}(x,y)  &  =\left\{  \min\left(  \frac{\gamma(x|y)}{p(x)},1\right)
\min\left(  \frac{\gamma(y|x)}{p(y)},1\right)  \right\}  ^{-1}\\
&  \geq\left\{  \min\left(  \frac{\gamma(x|y)}{p(x)},\frac{\gamma(y|x)}%
{p(y)}\right)  \right\}  ^{-1}=\max\left(  \frac{p(x)}{\gamma(x|y)}%
,\frac{p(y)}{\gamma(y|x)}\right)  =H\left(  x,y\right)  .
\end{align*}
Set $k(x,y)=M_{s}(x,y)$, Algorithm $M$ yields the same acceptance probability
as the special form of $\alpha_{HA}(x,y)$ in Equation (\ref{4b}):%
\begin{align*}
\alpha_{M}(x,y)  &  =\frac{p(y)}{M_{s}(x,y)\gamma(y|x)}=\min\left(
\frac{\gamma(x|y)}{p(x)},1\right)  \min\left(  \frac{\gamma(y|x)}%
{p(y)},1\right)  \frac{p(y)}{\gamma(y|x)}\\
&  =\min\left(  \frac{\gamma(x|y)}{p(x)},1\right)  \min\left(  \frac
{p(y)}{\gamma(y|x)},1\right)  .
\end{align*}

Return to the general $s(\cdot,\cdot)$ satisfying the Hastings condition
(\ref{01}), we may also define the following symmetric function,
\begin{equation}
m_{s}(x,y)=s(x,y)\left(  \frac{\gamma(x|y)}{p(x)}+\frac{\gamma(y|x)}%
{p(y)}\right)  ^{-1}=s(x,y)\left(  1+\frac{p(x)}{\gamma(x|y)}\frac
{\gamma(y|x)}{p(y)}\right)  ^{-1}\frac{p(x)}{\gamma(x|y)}\leq\frac
{p(x)}{\gamma(x|y)}. \label{15}%
\end{equation}
Because of its symmetrical property, $m_{s}(x,y)\leq L(x,y)$. With
$k(x,y)=m_{s}(x,y)$ in Equation (\ref{11}), we obtain in Algorithm $M$ the
same acceptance probability as that defined by $s(x,y)$ in HA:
\begin{equation}
\alpha_{M}(x,y)=\frac{m_{s}(x,y)\gamma(x|y)}{p(x)}=s(x,y)\left(  1+\frac
{p(x)}{\gamma(x|y)}\frac{\gamma(y|x)}{p(y)}\right)  ^{-1}=\alpha_{HA}(x,y).
\label{16}%
\end{equation}

For example, if $s(x,y)$ takes the form of Equation (\ref{4c}), then Equation
(\ref{15}) yields%
\begin{align*}
m_{s}(x,y)  &  =\min\left(  \frac{\gamma(x|y)}{p(x)},1\right)  \min\left(
\frac{p(y)}{\gamma(y|x)},1\right)  \frac{p(x)}{\gamma(x|y)}\\
&  \leq\min\left(  \frac{p(x)}{\gamma(x|y)},\frac{p(y)}{\gamma(y|x)}\right)
=L\left(  x,y\right)  .
\end{align*}
When we set $k(x,y)=m_{s}(x,y)$, Algorithm $M$ yields the same acceptance
probability as the special form of $\alpha_{HA}(x,y)$ in Equation (\ref{4b}):%
\[
\alpha_{M}(x,y)=\frac{m_{s}(x,y)\gamma(x|y)}{p(x)}=\min\left(  \frac
{\gamma(x|y)}{p(x)},1\right)  \min\left(  \frac{p(y)}{\gamma(y|x)},1\right)
.
\]
In the next subsection the reverse is proven.

\subsubsection{Algorithm $M$ is a special case of HA}

Algorithm $M$ is a special case of HA if, for any acceptance probability
$\alpha_{M}(\cdot,\cdot)$ in Algorithm $M$ (expressed in terms of
$k(\cdot,\cdot)$), we can find the same acceptance probability $\alpha
_{HA}(\cdot,\cdot)$ in HA.

\begin{description}
\item[Case 1:] When $k(x,y)\geq H(x,y)$: We set,%
\begin{align*}
s(x,y)  &  =\frac{1}{k(x,y)}\left(  \frac{p(x)}{\gamma(x|y)}+\frac
{p(y)}{\gamma(y|x)}\right)  =\frac{p(y)}{k(x,y)\gamma(y|x)}\left(
1+\frac{p(x)}{\gamma(x|y)}\frac{\gamma(y|x)}{p(y)}\right) \\
&  \leq1+\frac{p(x)}{\gamma(x|y)}\frac{\gamma(y|x)}{p(y)}.
\end{align*}
Substituting this form of $s(x,y)$ into Equation (\ref{01}) we obtain in HA
the same acceptance probability as $\alpha_{M}(x,y)$ in Equation (\ref{11})
when $k(x,y)\geq H(x,y)$.

\item[Case 2:] When $L(x,y)<k(x,y)<H(x,y)$: Equation (\ref{07b}) gives
$\alpha_{M}(x,y)=\alpha_{MH}(x,y)$. We thus set $s(x,y)=s_{MH}(x,y)$, as
defined in Equation (\ref{02}), to obtain the same acceptance probability.

\item[Case 3:] When $k(x,y)\leq L(x,y)$: We define,%
\begin{align*}
s(x,y)  &  =k(x,y)\left(  \frac{\gamma(x|y)}{p(x)}+\frac{\gamma(y|x)}%
{p(y)}\right)  =\frac{k(x,y)\gamma(x|y)}{p(x)}\left(  1+\frac{p(x)}%
{\gamma(x|y)}\frac{\gamma(y|x)}{p(y)}\right) \\
&  \leq1+\frac{p(x)}{\gamma(x|y)}\frac{\gamma(y|x)}{p(y)}.
\end{align*}
Substituting this form of $s(x,y)$ into Equation (\ref{01}) we obtain in HA
the same acceptance probability as $\alpha_{M}(x,y)$ in Equation (\ref{11})
when $k(x,y)\leq L(x,y)$.
\end{description}

Because Algorithm $M$ is a special case of HA and HA is also a special case of
Algorithm $M$, they are equivalent. It is worth noting, however, that the
relationship between $s(\cdot,\cdot)$ and $k(\cdot,\cdot)$ is not one-to-one.
The set of all $k(x,y)>0$ available to construct $\alpha_{M}(x,y)$ is larger
than the set of all $s(x,y)>0$ available to construct $\alpha_{HA}(x,y)$,
because $s(x,y)$ must also satisfy the Hastings' condition in Equation
(\ref{01}). In fact, for every $s(x,y)$, there are at least two distinct
expressions for $k(x,y)$: $M_{s}(x,y)\geq H(x,y)$ as defined in Equation
(\ref{13}), and $m_{s}(x,y)\leq L(x,y)$, as defined in Equation (\ref{15}). As
shown in Equation (\ref{07b}), all functions $k(x,y)$ that satisfy
$L(x,y)<k(x,y)<H(x,y)$ may be mapped to $s_{MH}(x,y)$.

\subsection{Algorithm $M$ and the Stein Algorithm}

Stein (in Liu 2001, p.~112) proposed an algorithm similar to HA in which the
acceptance probability $\alpha_{ST}(x,y)$ is expressed in terms of a symmetric
function $\delta(\cdot,\cdot)$ such that,%
\begin{equation}
0\leq\alpha_{ST}(x,y)=\frac{\delta(x,y)}{p(x)\gamma(y|x)}\leq1. \label{17}%
\end{equation}
By the same logic with which we showed the equivalence of Algorithm $M$ and
HA, we can show the equivalence of Algorithm $M$ and the Stein algorithm.

\subsubsection{The Stein algorithm is a special case of Algorithm $M$}

For each acceptance probability $\alpha_{ST}(\cdot,\cdot)$, expressed in terms
of $\delta(\cdot,\cdot)$, we can find the same acceptance probability
$\alpha_{M}(\cdot,\cdot)$: We define the symmetric function,%
\[
M_{\delta}(x,y)=\frac{p(x)p(y)}{\delta(x,y)}\geq\frac{p(x)p(y)}{p(x)\gamma
(y|x)}=\frac{p(y)}{\gamma(y|x)}.
\]
By symmetry, $M_{\delta}(x,y)\geq H(x,y)$. Then with $k(x,y)=M_{\delta}(x,y)$
in Equation (\ref{11}) we obtain in Algorithm $M$ the same acceptance
probability as that defined by $\delta(x,y)$ in the Stein algorithm:%
\[
\alpha_{M}(x,y)=\frac{\delta(x,y)}{p(x)p(y)}\frac{p(y)}{\gamma(y|x)}%
=\frac{\delta(x,y)}{p(x)\gamma(y|x)}=\alpha_{ST}(x,y).
\]

Alternatively, we can define the symmetric function,%
\[
m_{\delta}(x,y)=\frac{\delta(x,y)}{\gamma(x|y)\gamma(y|x)}\leq\frac
{p(x)\gamma(y|x)}{\gamma(x|y)\gamma(y|x)}=\frac{p(x)}{\gamma(x|y)}.
\]
By symmetry, $m_{\delta}(x,y)\leq L(x,y)$. Then with $k(x,y)=m_{\delta}(x,y)$
in Equation (\ref{11}) we obtain in Algorithm $M$ the same acceptance
probability as that defined by $s(x,y)$ in the Stein algorithm:%
\[
\alpha_{M}(x,y)=\frac{\delta(x,y)}{\gamma(x|y)\gamma(y|x)}\frac{\gamma
(x|y)}{p(x)}=\frac{\delta(x,y)}{p(x)\gamma(y|x)}=\alpha_{ST}(x,y).
\]

\subsubsection{Algorithm $M$ is a special case of the Stein algorithm}

For each acceptance probability $\alpha_{M}(\cdot,\cdot)$ expressed in terms
of $k(\cdot,\cdot)$, we can find the same acceptance probability $\alpha
_{ST}(\cdot,\cdot)$:

\begin{description}
\item[Case 1:] When $k(x,y)\geq H(x,y)$: We set%
\begin{align*}
\delta(x,y)  &  =\frac{p(x)p(y)}{k(x,y)}\leq p(x)p(y)\left(  \max\left\{
\frac{p(x)}{\gamma(x|y)},\frac{p(y)}{\gamma(y|x)}\right\}  \right)  ^{-1}\\
&  =p(x)p(y)\min\left\{  \frac{\gamma(x|y)}{p(x)},\frac{\gamma(y|x)}%
{p(y)}\right\} \\
&  =\min\left\{  p(y)\gamma(x|y),p(x)\gamma(y|x)\right\}  \leq p(x)\gamma
(y|x).
\end{align*}
Substituting this form of $\delta(x,y)$ into Equation (\ref{17}) we obtain in
the Stein algorithm the same acceptance probability as $\alpha_{M}(x,y)$ in
Equation (\ref{11}) when $k(x,y)\geq H(x,y)$.

\item[Case 2:] When $L(x,y)<k(x,y)<H(x,y)$: Due to Equation (\ref{07b}), we
set
\[
\delta(x,y)=\min\left\{  p(y)\gamma(x|y),p(x)\gamma(y|x)\right\}  \leq
p(x)\gamma(y|x),
\]
to obtain the same acceptance probability:%
\[
\alpha_{ST}(x,y)=\frac{\min\left\{  p(y)\gamma(x|y),p(x)\gamma(y|x)\right\}
}{p(x)\gamma(y|x)}=\alpha_{MH}(x,y).
\]

\item[Case 3:] When $k(x,y)\leq L(x,y)$: We set%
\begin{align*}
\delta(x,y)  &  =k(x,y)\gamma(x|y)\gamma(y|x)\leq\gamma(x|y)\gamma
(y|x)\min\left\{  \frac{p(x)}{\gamma(x|y)},\frac{p(y)}{\gamma(y|x)}\right\} \\
&  =\min\left\{  p(x)\gamma(y|x),p(y)\gamma(x|y)\right\}  \leq p(x)\gamma(y|x)
\end{align*}
Substituting this form of $\delta(x,y)$ into Equation (\ref{17}) we obtain in
the Stein algorithm the same acceptance probability as $\alpha_{M}(x,y)$ in
Equation (\ref{11}) when $k(x,y)\leq L(x,y)$.
\end{description}

Previously, the relationship between the Stein algorithm and HA was unclear.
Now we have shown that Algorithm $M$, the Stein algorithm and HA are all equivalent.

If Algorithm $M$ is equivalent to HA, then why do we introduce it? For the
rest of this paper, we will show how Algorithm $M$ may be developed
intuitively; we do not merely have to accept it because it satisfies detailed
balance. In the following section, we describe how Algorithm $M$ may be
obtained from a series of incremental modifications to the
Acceptance-Rejection (AR) algorithm (von Neumann, 1951).

\section{Markovian Acceptance-Rejection (MAR)}

\subsection{Acceptance-Rejection (AR)}

AR is a well-known algorithm that uses a proposal density $\gamma(\cdot)$ to
generate a sequence of independent variates from $p(\cdot)$. It requires a
\textquotedblleft majorizing coefficient\textquotedblright\ $M$ such that the
\textquotedblleft majorizing function\textquotedblright\ $M\gamma(\cdot)$
satisfies $M\gamma(z)\geq p(z)$ for all $z\in E$. Given $X_{n}=x\sim\pi
(\cdot)$, then $X_{n+1}\sim\pi(\cdot)$ can be generated by\bigskip

\textbf{Algorithm }$\mathbf{AR}$ (Acceptance-Rejection)

\begin{enumerate}
\item[AR1.] set $reject=1$

\item[AR2.] while $reject=1$

\begin{enumerate}
\item[AR2a.] generate $y\sim\gamma( \cdot) $ and $r\sim U( 0,1) $

\item[AR2b.] if $r\leq\frac{p(y)}{M\gamma(y)}\leq1$, output $X_{n+1}=y$, set
$reject=0$
\end{enumerate}

\item[AR3.] endwhile\bigskip
\end{enumerate}

AR is easy to understand conceptually. The condition $M\gamma(\cdot)\geq
p(\cdot)$ assures that the surface $M\gamma(\cdot)$ is above that of
$p(\cdot)$. With $r\sim U(0,1),$ every pair $(y\sim\gamma(\cdot),rM\gamma(y))$
is uniformly distributed under the surface $M\gamma(\cdot)$. Of these, those
that satisfy the condition in Step AR2b (and hence are accepted) are also
uniformly distributed under the surface $p(\cdot)$. These $y$ variates have
density $\pi(\cdot)$. (See Minh, 2001, Chap.~13.)

\subsection{Independence\ Markovian Acceptance-Rejection (IMAR)}

We now present a simple modification of AR into what we call the
\textquotedblleft Independence Markovian
Acceptance-Rejection\textquotedblright\ algorithm (IMAR). Given $X_{n}%
=x\sim\pi(\cdot),$ and a proposed density $\gamma(\cdot)$ (which is
independent of $x$), $X_{n+1}\sim\pi(\cdot)$ can be generated by\bigskip

\textbf{Algorithm }$\mathbf{IMAR}$ (Independence Markovian Acceptance-Rejection)

\begin{enumerate}
\item[IMA1.] generate $y\sim\gamma(\cdot)$ and $r\sim U(0,1)$

\item[IMA2.] if $r\leq\alpha_{IMA}(x,y)=\frac{p(y)}{M\gamma(y)}\leq1$, output
$X_{n+1}=y$

\item[IMA3.] else, output $X_{n+1}=x$\bigskip
\end{enumerate}

The main distinction between AR and IMAR is that, when a proposed variate $y$
is rejected in IMAR, the variate $x$ is repeated. Suppose that a sequence of
proposed variates is $y_{1},y_{2},y_{3},y_{4},y_{5},y_{6},...$. If AR accepts
$y_{1},y_{3},y_{6,}...$, then with the same sequence of random numbers, IMAR
would generate $y_{1},y_{1},y_{3},y_{3},y_{3},y_{6},...$. Because of the
repetitions, IMAR does not generate independent variates; rather it is a
Markov Chain Monte Carlo method that satisfies detailed balance with respect
to $p(\cdot)$: For all $x,y\in E$ and $x\neq y$,
\[
p(x)P_{IMA}(y|x)=p(x)\frac{p(y)}{M\gamma(y)}\gamma(y)=P_{IMA}(x|y)p(y).
\]

As in AR, the expected number of times that $z$ is delivered in Step IMA2 in a
simulation is proportional to $p(z)$. Also, the expected number of
duplications in Step IMA3 is the same for all variates, which is $M-1$, the
expected number of consecutive rejections in the corresponding AR. Thus the
expected total number of times that $z$ and its duplicates are delivered is
proportional to $Mp(z)$, or to $\pi(z)$ because $M$ is a constant.

While it is hard to find a majorizing coefficient\ $M$ such that
$M\gamma(z)\geq p(z)$ for all $z\in E$, it is easier to find a
\textquotedblleft deficient\textquotedblright\ majorizing coefficient $M$ such
that $M\gamma(z)\geq p(z)$ for some $z\in E$. In this case, it is well known
that AR produces variates from $\min\left\{  p(\cdot),M\gamma(\cdot)\right\}
$. This is also true for IMAR, in which $M\gamma(\cdot)$ serves as the
majorizing function for $\min\left\{  p(\cdot),M\gamma(\cdot)\right\}  $,
resulting in the acceptance probability%
\[
\alpha_{D}(x,y)=\frac{\min\left\{  p(y),M\gamma(y)\right\}  }{M\gamma(y)}%
=\min\left\{  \frac{p(y)}{M\gamma(y)},1\right\}  .
\]
For future reference, it is important to note that, even with a deficient
majorizing constant, AR and IMAR still generate variates $y\sim\pi(\cdot)$
within the region $\left\{  z:M\gamma(z)\geq p(z)\right\}  $.

\subsection{Markovian Acceptance-Rejection (MAR)}

As a generalization of IMAR, we now allow the proposal density $\gamma(
\cdot|x) $ to be dependent on the chain's current value $X_{n}=x$.

If we knew beforehand that AR accepts $y_{3}$ out of 3 proposed variates
$y_{1},y_{2}$ and $y_{3}$, then all we would need is a majorizing coefficient
$M$ such that $M\gamma(y_{i})\geq p(y_{i})$ for $i=1,2,3$. The problem is that
the number of consecutive rejections before an acceptance in AR may be
infinite, and $y$ can be anywhere in $E$. Furthermore, to generate independent
variates, the majorizing coefficient $M$ in AR must be independent of the
current variate $x$. So AR needs an \textquotedblleft
absolute\textquotedblright\ majorizing coefficient $M$\ such that
$M\gamma(z)\geq p(z)$ for all $z\in E$.

When the proposal density $\gamma(\cdot|x)$ is allowed to be dependent on
$X_{n}=x$, the requirement of an absolute majorizing coefficient $M$, however,
is too restrictive: if there is a pair $\eta,\xi\in E$ such that $\gamma
(\eta|\xi)=0$ and $p(\eta)>0$, then we must have $M=\infty$.

Fortunately, similar to IMAR, in the following Algorithm MAR, which allows
$\gamma(\cdot|x)$ to be dependent on $x$, either the current variate $x$ or
the proposed variate $y$ must be delivered in each iteration. So, instead of
requiring an absolute majorizing coefficient $M$, we only need a
\textquotedblleft relative\textquotedblright\ majorizing coefficient
$M(\cdot,\cdot)>0$ that may change with each pair $(x,y)$, so long as, for all
$x,y\in E$, $M(x,y)\gamma(x|y)\geq p(x)$ and $M(x,y)\gamma(y|x)\geq p(y)$,
or,
\begin{equation}
M(x,y)\geq\max\left\{  \frac{p(x)}{\gamma(x|y)},\frac{p(y)}{\gamma
(y|x)}\right\}  =H(x,y). \label{20}%
\end{equation}

It is necessary that the relative\ majorizing coefficient $M(\cdot,\cdot)$ is
symmetric, in order to preserve the balance of flows from $x$ to $y$ and from
$y$ to $x$. We thus may write $M(\cdot,\cdot)$ in terms of any symmetric
function $C(\cdot,\cdot)\geq1$:%
\begin{equation}
M(x,y)=C(x,y)\max\left\{  \frac{p(x)}{\gamma(x|y)},\frac{p(y)}{\gamma
(y|x)}\right\}  . \label{21}%
\end{equation}

Given $X_{n}=x\sim\pi(\cdot)$, the following algorithm may be used to generate
$X_{n+1}\sim\pi(\cdot)$:\bigskip

\textbf{Algorithm }$\mathbf{MAR}$ (Markovian Acceptance-Rejection)

\begin{enumerate}
\item[MA1.] generate $y\sim\gamma(\cdot|x)$ and $r\sim U(0,1)$

\item[MA2.] if $r\leq\alpha_{MA}(x,y)$, output $X_{n+1}=y$

\item[MA3.] else, output $X_{n+1}=x$\bigskip
\end{enumerate}

\noindent where, with $M\left(  x,y\right)  $ defined in Equation (\ref{21}),
\begin{align}
\alpha_{MA}(x,y)  &  =\frac{p(y)}{M(x,y)\gamma(y|x)}\label{18}\\
&  =\frac{p(y)}{C(x,y)\max\left\{  \frac{p(x)}{\gamma(x|y)},\frac{p(y)}%
{\gamma(y|x)}\right\}  \gamma(y|x)}\label{22a}\\
&  =\frac{1}{C(x,y)}\min\left\{  \frac{\gamma(x|y)}{p(x)}\frac{p(y)}%
{\gamma(y|x)},1\right\}  \leq1. \label{22}%
\end{align}
\bigskip

As with IMAR, it is straightforward to show that, if $M(\cdot,\cdot)$ is
symmetric, then the transition kernel MAR satisfies detailed balance with
respect to $p(\cdot)$: For all $x,y\in E$ and $x\neq y$,%
\[
p(x)P_{MA}(y|x)=p(x)\frac{p(y)}{M(x,y)\gamma(y|x)}\gamma(y|x)=P_{MA}%
(x|y)p(y).
\]

As previously noted, using a deficient (absolute) majorizing coefficient
$M$\ in IMAR still generates variates $y\sim p(\cdot)$ within the region
$\left\{  z:M\gamma(z)\geq p(z)\right\}  $. A relative majorizing coefficient
may be a deficient (absolute) majorizing coefficient,\ but is sufficient for
$x$ and $y$, because both $x$ and $y$ are within the region $\left\{
z:M(x,z)\gamma(z|x)\geq p(z)\right\}  $.

We now show that BK and MH are two special cases of MAR.

\subsection{BK in MAR}

As a special case of MAR, we let
\begin{equation}
C(x,y)=\left(  \frac{p(y)}{\gamma(y|x)}+\frac{p(x)}{\gamma(x|y)}\right)
\left(  \max\left\{  \frac{p(x)}{\gamma(x|y)},\frac{p(y)}{\gamma
(y|x)}\right\}  \right)  ^{-1}>1. \label{33}%
\end{equation}
Then the acceptance probability in Equation (\ref{22a}) becomes the Barker's
acceptance probability $\alpha_{BK}(x,y)$ in Equation (\ref{04}):
\[
\alpha_{MA}(x,y)=\frac{p(y)}{\left(  \frac{p(y)}{\gamma(y|x)}+\frac
{p(x)}{\gamma(x|y)}\right)  \gamma(y|x)}=\left(  1+\frac{p(x)}{\gamma
(x|y)}\frac{\gamma(y|x)}{p(y)}\right)  ^{-1}=\alpha_{BK}(x,y).
\]

\subsection{MH in MAR}

If we set $C(x,y)=1$, the acceptance probability $\alpha_{MA}(x,y)$ in
Equation (\ref{22}) become the acceptance probability $\alpha_{MH}(x,y)$ in
Equation (\ref{03}) and MAR becomes MH.

Peskun (1973) introduced partial ordering on transition kernels to prove that,
with the same proposal densities, $\alpha_{MH}(x,y)$ is optimal in terms of
minimizing the asymptotic variance of sample path averages. In the MAR
framework, it is straightforward to see that the acceptance probability in
Equation (\ref{22}) is maximized when $C(x,y)=1$. Thus MH has the highest
acceptance probability of all Hastings algorithms.

Conceptually, the most efficient majorizing function $M\gamma(\cdot)$ in AR is
the one that \textquotedblleft touches\textquotedblright\ the target density
$p(\cdot)$ at one point. Similarly, when $C(x,y)=1$, Equation (\ref{21}) shows
that either $M(x,y)\gamma(x|y)=p(x)$ or $M(x,y)\gamma(y|x)=p(y)$. Any higher
value of $\ C(x,y)$ only results in unnecessarily rejecting some proposed
variates. This is what happens in BK, where $C_{BK}(x,y)>1$ as in Equation
(\ref{33}).

We have derived and explained MAR intuitively. It turns out that MAR\ is
equivalent to Algorithm $M$.

\subsection{MAR\ and Algorithm $M$}

MAR is a special case of Algorithm $M$ if, for any acceptance probability
$\alpha_{MA}(\cdot,\cdot)$ (defined in terms of the relative\ majorizing
coefficient $M(x,y)$) in MAR, we can find the same acceptance probability
$\alpha_{M}(\cdot,\cdot)$ in Algorithm $M$. We achieve this simply by letting
$k(x,y)=M(x,y)\geq H(x,y)$ in Equation (\ref{11}), resulting in $\alpha
_{M}(x,y)=\alpha_{MA}(x,y)$.

For equivalence, the reverse must also be true; that is, Algorithm $M$ is a
special case of MAR. We now show that, for any acceptance probability
$\alpha_{M}(x,y)$ (defined in terms of $k(x,y)$), we can also find the same
$\alpha_{MA}(x,y)$ in MAR. Consider%
\[
M_{k}(x,y)=k(x,y)\max\left\{  \frac{p(x)}{k(x,y)\gamma(x|y)},1\right\}
\max\left\{  \frac{p(y)}{k(x,y)\gamma(y|x)},1\right\}  .
\]

$M_{k}(x,y)$ is a relative majorizing coefficient because it satisfies the
inequality (\ref{20}):

\begin{description}
\item[Case 1.] When $k(x,y)\geq H(x,y)$:
\[
M_{k}(x,y)=k(x,y)\geq H(x,y)
\]

\item[Case 2.] When $L(x,y)<k(x,y)<H(x,y)$:%
\begin{align}
M_{k}(x,y)  &  =\left\{
\begin{array}
[c]{ll}%
\frac{p(y)}{\gamma(y|x)} & \text{if }\frac{p(x)}{\gamma(x|y)}\leq
k(x,y)\leq\frac{p(y)}{\gamma(y|x)}\\
\frac{p(x)}{\gamma(x|y)} & \text{if }\frac{p(y)}{\gamma(y|x)}\leq
k(x,y)\leq\frac{p(x)}{\gamma(x|y)}%
\end{array}
\right. \label{24}\\
&  =\max\left\{  \frac{p(x)}{\gamma(x|y)},\frac{p(y)}{\gamma(y|x)}\right\}
=H(x,y)\nonumber
\end{align}

\item[Case 3.] If $k(x,y)\leq L(x,y)$:
\begin{align}
M_{k}(x,y)  &  =\frac{1}{k(x,y)}\frac{p(x)}{\gamma(x|y)}\frac{p(y)}%
{\gamma(y|x)}\label{25a}\\
&  \geq\frac{1}{\min\left\{  \frac{p(x)}{\gamma(x|y)},\frac{p(y)}{\gamma
(y|x)}\right\}  }\frac{p(x)}{\gamma(x|y)}\frac{p(y)}{\gamma(y|x)}=\max\left\{
\frac{p(x)}{\gamma(x|y)},\frac{p(y)}{\gamma(y|x)}\right\}  =H(x,y)\nonumber
\end{align}

\end{description}

Setting $M(x,y)=M_{k}(x,y)$ in Equation (\ref{18}) yields:%
\begin{align*}
\alpha_{MA}(x,y)  &  =\frac{p(y)}{k(x,y)\max\left\{  \frac{p(x)}%
{k(x,y)\gamma(x|y)},1\right\}  \max\left\{  \frac{p(y)}{k(x,y)\gamma
(y|x)},1\right\}  \gamma(y|x)}\\
&  =\min\left\{  \frac{k(x,y)\gamma(x|y)}{p(x)},1\right\}  \min\left\{
\frac{k(x,y)\gamma(y|x)}{p(y)},1\right\}  \frac{p(y)}{k(x,y)\gamma(y|x)}\\
&  =\min\left\{  \frac{k(x,y)\gamma(x|y)}{p(x)},1\right\}  \min\left\{
\frac{p(y)}{k(x,y)\gamma(y|x)},1\right\}  =\alpha_{M}(x,y)
\end{align*}

Hence Algorithm $M$ and MAR\ are equivalent.

\subsection{MAR\ and HA}

Because MAR\ is equivalent to Algorithm $M$, and Algorithm $M$ is equivalent
to HA, MAR and HA are equivalent. To show this directly, for every
$\alpha_{MA}(x,y)$ defined by $M(x,y)\geq H(x,y)$ in MAR, we set,
\begin{align}
s(x,y)  &  =\frac{1}{M(x,y)}\left(  \frac{p(x)}{\gamma(x|y)}+\frac
{p(y)}{\gamma(y|x)}\right)  =\frac{p(y)}{M(x,y)\gamma(y|x)}\left(
1+\frac{p(x)}{\gamma(x|y)}\frac{\gamma(y|x)}{p(y)}\right) \label{26a}\\
&  \leq1+\frac{p(x)}{\gamma(x|y)}\frac{\gamma(y|x)}{p(y)}.\nonumber
\end{align}
Substituting this form of $s(x,y)$ into Equation (\ref{01}) we obtain in HA
the same acceptance probability $\alpha_{MA}(x,y)$. Hence MAR is a special
case of HA. On the other hand, for each $\alpha_{HA}(x,y)$ defined by
$s(x,y)$, we set $M(x,y)=M_{s}(x,y)\geq H(x,y)$, where $M_{s}(x,y)$ was
defined in Equation (\ref{13}). Like Equation (\ref{14}), Equation (\ref{18})
then yields in MAR the same acceptance probability as $\alpha_{HA}(x,y)$ in
HA. Thus, HA is a special case of MAR. Therefore, MAR\ and HA are equivalent.

Equations (\ref{13}) and (\ref{26a}) show that there is a one-to-one mapping
between the set of all symmetric functions $s(x,y)$ satisfying Hastings'
condition, Equation (\ref{01}), and the set of all symmetric functions
$M(x,y)$ in the form of Equation (\ref{21}). However, unlike the mysterious
$s(\cdot,\cdot)$, $M(\cdot,\cdot)$ has a very intuitive interpretation of
being a relative majorizing coefficient.

Thus far, we have intuitively derived HA as MAR, which is Algorithm $M$ in
which $k( \cdot,\cdot) $ is sufficiently large to be a relative majorizing
coefficient. We now show that HA can also be explained in terms of an
algorithm \textquotedblleft dual\textquotedblright\ to MAR, which is Algorithm
$M$ with a sufficiently small coefficient $k( \cdot,\cdot) $.

\section{Markovian Minorizing (MIR)}

\subsection{Independence Markovian Minorizing (IMIR)}

We now return to the assumption that the proposal densities are independent of
the chain's current variate, or $\gamma(\cdot|x)=\gamma(\cdot)$. We also
assume that the support of $\gamma\left(  \cdot\right)  $\ includes that of
$p\left(  \cdot\right)  $\ and there is an \textquotedblleft absolute
minorizing coefficient\textquotedblright\ $m$ such that $m\gamma(z)\leq p(z)$
for all $z\in E$.

Consider the following algorithm that we call the \textquotedblleft
Independence Markovian Minorizing\textquotedblright\ algorithm (IMIR): Given
$X_{n}=x\sim\pi(\cdot)$, then $X_{n+1}\sim\pi(\cdot)$ can be generated
by\bigskip

\textbf{Algorithm }$\mathbf{IMIR}$\textbf{ }(Independence Markovian Minorizing)

\begin{enumerate}
\item[IMI1.] generate $y\sim\gamma(\cdot)$ and $r\sim U(0,1)$

\item[IMI2.] if $r\leq\alpha_{IMI}(x,y)=\frac{m\gamma(x)}{p(x)}\leq1$, output
$X_{n+1}=y$

\item[IMI3.] else, output $X_{n+1}=x$\bigskip
\end{enumerate}

The transition kernel of this algorithm is $P_{IMI}(y|x)=\alpha_{IMI}%
(x,y)\gamma(y)$ for all $x,y\in E$ and $x\neq y$, which satisfies detailed
balance with respect to $p(\cdot)$:%
\begin{equation}
p(x)P_{IMI}(y|x)=p(x)\frac{m\gamma(x)}{p(x)}\gamma(y)=P_{IMI}%
(x|y)p(y).\nonumber
\end{equation}

We may not have an absolute minorizing coefficient $m$, but only a
\textquotedblleft deficient\textquotedblright\ minorizing coefficient $m$ such
that $m\gamma(z)\leq p(z)$ for some $z\in E$. Using $m\gamma(\cdot)$ as the
minorizing function for $\max\left\{  p(\cdot),m\gamma(\cdot)\right\}  $ in
Algorithm IMIR, the acceptance probability $\alpha_{IMI}(x,y)$ becomes%
\[
\alpha_{d}(x,y)=\frac{m\gamma(x)}{\max\left\{  p(x),m\gamma(x)\right\}  }%
=\min\left\{  \frac{m\gamma(x)}{p(x)},1\right\}  ,
\]
and Algorithm IMIR generates variates from $\max\left\{  p(\cdot
),m\gamma(\cdot)\right\}  $. Note that, similar to the discussion for IMAR,
even with a deficient minorizing constant, IMIR still generates variates
$y\sim\pi(\cdot)$ within the region $\left\{  z:m\gamma(z)\leq p(z)\right\}  $.

We wrote Algorithm IMIR in the form consistent with that of all other
algorithms in this paper. However, we do not need to generate $y\sim
\gamma(\cdot)$ in Step IMI1 if $r>\alpha_{IMI}(x,y)$ in Step IMI2. For a more
intuitive understanding, Algorithm IMIR can also be written as,\bigskip

\bigskip\textbf{Algorithm }$\mathbf{IMJ}$

\begin{enumerate}
\item[IMJ1.] generate $r\sim U(0,1)$

\item[IMJ2.] if $r\leq\alpha_{IMI}(x,y)=\frac{m\gamma(x)}{p(x)}\leq1$,
generate $y\sim\gamma(\cdot)$, output $X_{n+1}=y$

\item[IMJ3.] else, output $X_{n+1}=x$\bigskip
\end{enumerate}

In a simulation, the expected number of times that $x=y\sim\gamma(\cdot)$ is
delivered in Step IMJ2 is proportional to $\gamma(x)$. Furthermore, for each
$x$ so delivered, it is duplicated until the first success in a sequence of
Bernoulli trials with success probability $m\gamma(x)/p(x)$; the expected
number of its duplications is $p(x)/\left[  m\gamma(x)\right]  \geq1$. Thus in
a simulation, the expected total number of times that $x$ is delivered is
proportional to $\gamma(x)\left\{  p(x)/\left[  m\gamma(x)\right]  \right\}
=p(x)/m$, or to $p(x)$ because $m$ is a constant.

In Minh et al (2012) we used the minorizing coefficient $m$ to make any Markov
Chain Monte Carlo method regenerative.

\subsection{Markovian Minorizing (MIR)}

If the proposal density is dependent on $x$, taking the form $\gamma(\cdot
|x)$, the requirement of an \textquotedblleft absolute\textquotedblright%
\ minorizing coefficient $m$ such that $m\gamma(y|x)\leq p(y)$ for all $x,y\in
E$ is too restrictive, and often can only be satisfied when $m=0$.
Fortunately, similar to MAR, given the current variate $x$ and proposed
variate $y$, there is no need for such an absolute minorizing coefficient, but
only a \textquotedblleft relative\textquotedblright\ minorizing
coefficient\ $m(x,y)>0$ such that $m(x,y)\gamma(x|y)\leq p(x)$.

It is important that $m(x,y)$ is symmetric, as it preserves the balance of
flows from $x$ to $y$ and from $y$ to $x$. Therefore $m(x,y)$ must be such
that%
\begin{equation}
m(x,y)\leq\min\left\{  \frac{p(x)}{\gamma(x|y)},\frac{p(y)}{\gamma
(y|x)}\right\}  =L(x,y). \label{27}%
\end{equation}

With the previously defined symmetric function $C(\cdot,\cdot)$ such that
$C(\cdot,\cdot)\geq1$, $m(x,y)$ can be written in the following form:%
\begin{equation}
m(x,y)=\frac{1}{C(x,y)}\min\left\{  \frac{p(x)}{\gamma(x|y)},\frac
{p(y)}{\gamma(y|x)}\right\}  . \label{28}%
\end{equation}

Given $X_{n}=x\sim p(\cdot)$, the following \textquotedblleft Markovian
Minorizing\textquotedblright\ algorithm (MIR) may be used to generate
$X_{n+1}\sim\pi(\cdot)$:\bigskip

\textbf{Algorithm }$\mathbf{MIR}$\textbf{ }(Markovian Minorizing)\textbf{:}

\begin{enumerate}
\item[MI1.] generate $y\sim\gamma(\cdot|x)$ and $r\sim U(0,1)$

\item[MI2.] if $r\leq\alpha_{MI}(x,y)$, output $X_{n+1}=y$

\item[MI3.] else, output $X_{n+1}=x$\bigskip
\end{enumerate}

\noindent where%
\begin{equation}
\alpha_{MI}(x,y)=\frac{m(x,y)\gamma(x|y)}{p(x)}=\frac{1}{C(x,y)}\min\left\{
\frac{\gamma(x|y)}{p(x)}\frac{p(y)}{\gamma(y|x)},1\right\}  \leq1. \label{39}%
\end{equation}

The transition kernel of Algorithm MIR is $P_{MI}(y|x)=\alpha_{MI}%
(x,y)\gamma(y|x)$ for all $x,y\in E$ and $x\neq y$, which satisfies detailed
balance with respect to $p(\cdot)$:%
\[
p(x)P_{MI}(y|x)=p(x)\frac{m(x,y)\gamma(x|y)}{p(x)}\gamma(y|x)=P_{MI}%
(x|y)p(y).
\]

As previously noted, IMIR with a deficient (absolute) minorizing coefficient
$m$ generates variates $y\sim p(\cdot)$ within the region $\left\{
z:m\gamma(z)\leq p(z)\right\}  $. Similarly, the relative minorizing
coefficient $m(x,y)$ in the form of (\ref{28}) may be deficient as an absolute
minorizing coefficient, but it was chosen so that both $x$ and $y$ are in the
region $\left\{  z:m(x,z)\gamma(z|x)\leq p(z)\right\}  $.

\subsection{MIR\ and HA}

MIR\ and MAR are equivalent because the acceptance probability of MIR in
Equation (\ref{39}) is identical with that of MAR in Equation (\ref{22}). MIR
therefore is also equivalent to HA. In fact, for any $\alpha_{MI}(x,y)$
defined by $m(x,y)$ in MIR, we set%
\begin{align}
s(x,y)  &  =m(x,y)\left(  \frac{\gamma(x|y)}{p(x)}+\frac{\gamma(y|x)}%
{p(y)}\right)  =\frac{m(x,y)\gamma(x|y)}{p(x)}\left(  1+\frac{p(x)}%
{\gamma(x|y)}\frac{\gamma(y|x)}{p(y)}\right) \label{40}\\
&  \leq1+\frac{p(x)}{\gamma(x|y)}\frac{\gamma(y|x)}{p(y)}.\nonumber
\end{align}
Substituting this form of $s(x,y)$ into Equation (\ref{01}) we obtain the same
acceptance probability as $\alpha_{MI}(x,y)$ in HA . Conversely, for any
$s(x,y)$ that defines $\alpha_{HA}(x,y)$, we let $m(x,y)=m_{s}(x,y)\leq
L(x,y)$ as defined in Equation (\ref{15}). Then, similar to Equation
(\ref{16}), Equation (\ref{39}) yields $\alpha_{MI}(x,y)=\alpha_{HA}(x,y)$.

We have derived HA as MIR. Equations (\ref{15}) and (\ref{40}) show that there
is a one-to-one mapping between the set of all symmetric functions $s(x,y)$
satisfying Hastings' condition (\ref{01}) and the set of all symmetric
functions $m(x,y)$ satisfying condition (\ref{27}). However, $m(\cdot,\cdot)$
has a very intuitive interpretation of being the relative minorizing coefficients.

\subsection{MIR and Algorithm $M$}

Replacing $k(x,y)$ with $m(x,y)\leq L(x,y)$ in Equation (\ref{11}), we obtain
$\alpha_{M}(x,y)=\alpha_{MI}(x,y)$. MIR therefore is a special case of
Algorithm $M$ in which $k(x,y)$ is low enough to be a relative minorizing
coefficient. The reverse is also true; that is, for every $k(x,y)>0$, we
define%
\[
m_{k}(x,y)=k(x,y)\min\left\{  \frac{p(x)}{k(x,y)\gamma(x|y)},1\right\}
\min\left\{  \frac{p(y)}{k(x,y)\gamma(y|x)},1\right\}  ,
\]
which is a relative minorizing coefficient because it satisfies the inequality
(\ref{27}):

\begin{description}
\item[Case 1.] When $k(x,y)\geq H(x,y)$:
\begin{align}
m_{k}(x,y)  &  =\frac{1}{k(x,y)}\frac{p(x)}{\gamma(x|y)}\frac{p(y)}%
{\gamma(y|x)}\label{32}\\
&  \leq\frac{1}{\max\left\{  \frac{p(x)}{\gamma(x|y)},\frac{p(y)}{\gamma
(y|x)}\right\}  }\frac{p(x)}{\gamma(x|y)}\frac{p(y)}{\gamma(y|x)}\nonumber\\
&  =\min\left\{  \frac{p(x)}{\gamma(x|y)},\frac{p(y)}{\gamma(y|x)}\right\}
=L(x,y)\nonumber
\end{align}

\item[Case 2.] When $L(x,y)<k(x,y)<H(x,y)$:%
\begin{align}
m_{k}(x,y)  &  =\left\{
\begin{array}
[c]{ll}%
\frac{p(x)}{\gamma(x|y)} & \text{if }\frac{p(x)}{\gamma(x|y)}\leq
k(x,y)\leq\frac{p(y)}{\gamma(y|x)}\\
\frac{p(y)}{\gamma(y|x)} & \text{if }\frac{p(y)}{\gamma(y|x)}\leq
k(x,y)\leq\frac{p(x)}{\gamma(x|y)}%
\end{array}
\right. \nonumber\\
&  =\min\left\{  \frac{p(x)}{\gamma(x|y)},\frac{p(y)}{\gamma(y|x)}\right\}
=L(x,y) \label{34}%
\end{align}

\item[Case 3.] When $k(x,y)\leq L(x,y)$:
\[
m_{k}(x,y)=k(x,y)\leq L(x,y)
\]

\end{description}

Letting $m(x,y)=$ $m_{k}(x,y)$ in Equation (\ref{39}) yields:%
\begin{align*}
\alpha_{MI}(x,y)  &  =\min\left\{  \frac{p(y)}{k(x,y)\gamma(y|x)},1\right\}
\min\left\{  \frac{p(x)}{k(x,y)\gamma(x|y)},1\right\}  \frac{k(x,y)\gamma
(x|y)}{p(x)}\\
&  =\min\left\{  \frac{p(y)}{k(x,y)\gamma(y|x)},1\right\}  \min\left\{
\frac{k(x,y)\gamma(x|y)}{p(x)},1\right\}  =\alpha_{M}(x,y).
\end{align*}
Hence Algorithm $M$ is also a special case of MIR. They are equivalent.

\section{Summary}

We now summarize the relationship between Algorithm $M$, MAR and MIR by
explaining what happens when $k(x,y)$ reduces from a very high value to a very
low one.

Before doing so, we write Algorithm $M$ in a two-stage form: Given
$X_{n}=x\sim\pi(\cdot)$, then $X_{n+1}\sim\pi(\cdot)$ can be generated
by\bigskip

\textbf{Algorithm }$\mathbf{L}$\textbf{:}

\begin{enumerate}
\item[L1.] generate $y\sim\gamma(\cdot|x)$ and $r_{1}\sim U(0,1)$

\item[L2.] if $r_{1}>\min\left\{  \frac{k(x,y)\gamma(x|y)}{p(x)},1\right\}  $,
output $X_{n+1}=x$ (MIR, type-$x$ duplication)

\item[L3.] else,

\begin{enumerate}
\item[L3a.] generate $r_{2}\sim U(0,1)$

\item[L3b.] if $r_{2}>\min\left\{  \frac{p(y)}{k(x,y)\gamma(y|x)},1\right\}
$, output $X_{n+1}=x$ (MAR, type-$y$ duplication)

\item[L3c.] else output $X_{n+1}=y$
\end{enumerate}

\item[L4.] endif\bigskip
\end{enumerate}

This allows us to classify the duplication of $x$ either as a
\textquotedblleft type-$x$\textquotedblright\ duplication, which occurs in
Step L2, or as a \textquotedblleft type-$y$\textquotedblright\ duplication,
which occurs in Step L3b. (The conditions in Steps L2 and L3b may be
switched.) The probability of a type-$x$ duplication is $1-\min\left\{
\frac{k(x,y)\gamma(x|y)}{p(x)},1\right\}  $ and the probability of a type-$y$
duplication is $\min\left\{  \frac{k(x,y)\gamma(x|y)}{p(x)},1\right\}  \left(
1-\min\left\{  \frac{p(y)}{k(x,y)\gamma(y|x)},1\right\}  \right)  $.\ Thus the
probability of $x$ being duplicated is%
\begin{align*}
&  1-\min\left\{  \frac{k(x,y)\gamma(x|y)}{p(x)},1\right\}  +\min\left\{
\frac{k(x,y)\gamma(x|y)}{p(x)},1\right\}  \left(  1-\min\left\{  \frac
{p(y)}{k(x,y)\gamma(y|x)},1\right\}  \right) \\
&  =1-\min\left\{  \frac{p(y)}{k(x,y)\gamma(y|x)},1\right\}  \min\left\{
\frac{k(x,y)\gamma(x|y)}{p(x)},1\right\}  ,
\end{align*}
which is the same as the probability of duplicating $x$ in Algorithm $M$.

\begin{description}
\item[Case 1.] When $k(x,y)\geq H(x,y)$: We start with a very high value of
$k(x,y)$ such that $k(x,y)\geq H(x,y)$. Then $k(x,y)$ is a relative majorizing
coefficient $M(x,y)$ and Algorithm $M$ is MAR, utilizing only type-$y$
duplications. There is a corresponding MIR with a relative minorizing
coefficient $m_{k}(x,y)$ as defined in Equation (\ref{32}), utilizing only
type-$x$ duplications. As $k(x,y)=M(x,y)$ decreases, both $\alpha
_{M}(x,y)=\alpha_{MA}(x,y)$ and $m_{k}(x,y)$ increase. When $k(x,y)=M(x,y)$
decreases to $H(x,y)$, $m_{k}(x,y)$ increases to $L(x,y)$ and the acceptance
probability $\alpha_{M}(x,y)=\alpha_{MA}(x,y)$ reaches its maximum value
$\alpha_{MH}(x,y)$.

\item[Case 2.] When $L(x,y)<k(x,y)<H(x,y)$: When $k(x,y)$ further decreases
below $H(x,y)$, it becomes \textquotedblleft too deficient\textquotedblright%
\ for MAR to generate variates from $p(\cdot)$ with type-$y$ duplications
alone; type-$x$ duplications are also needed to make Algorithm $M$ equivalent
to MH. As the value of $k(x,y)$ decreases further, we see fewer type-$y$
duplications and more type-$x$ duplications, but the acceptance probability
$\alpha_{M}(x,y)$ remains at its maximum value $\alpha_{MH}(x,y)$. In this
case, regardless of the value of $k(x,y)$, there is a corresponding relative
majorizing coefficient $M_{k}(x,y)=H(x,y)$ as in Equation (\ref{24}) and a
corresponding relative minorizing coefficient $m_{k}(x,y)=L(x,y)$ as in
Equation (\ref{34}).

\item[Case 3.] When $k(x,y)\leq L(x,y)$: Further decreasing $k(x,y)$ below
$L(x,y)$, we see Algorithm $M$ becomes MIR, utilizing only type-$x$
duplications, with $k(x,y)$ as a relative minorizing coefficient $m(x,y)$.
There is a corresponding MAR with a relative majorizing coefficient
$M_{k}(x,y)$ defined in Equation (\ref{25a}), utilizing only type-$y$
duplications. As $k(x,y)=m(x,y)$ decreases from $L(x,y)$, $M_{k}(x,y)$
increases from $H(x,y)$, and the acceptance probability $\alpha_{M}%
(x,y)=\alpha_{MI}(x,y)$ decreases from its maximum value $\alpha_{MH}(x,y)$.
\end{description}

Algorithm $M$ is a combination of MAR (which is HA), MIR (which is also HA)
and MH (which is the optimal case of HA). It is not more general than HA, but
it is easier to understand intuitively.

\section*{References}

\begin{enumerate}
\item Barker, A. A. (1965) \textquotedblleft Monte Carlo Calculations of the
Radial Distribution Functions for a Proton-electron Plasma,\textquotedblright%
\ \textit{Australian Journal of Physics,} 18, 119-33.

\item Billera, L. J. and Diaconis, P. (2001) \textquotedblleft A Geometric
Interpretation of the Metropolis-Hastings Algorithm,\textquotedblright%
\ \textit{Statistical Science}, 16, 335-339.

\item Chib, S. and Greenberg, E. (1995) \textquotedblleft Understanding the
Metropolis-Hastings Algorithm,\textquotedblright\ \textit{The American
Statistician}, 49, 327-335.

\item Dongarra, J.; Sullivan, F. (2000) \textquotedblleft Guest Editors'
Introduction: The Top 10 Algorithms,\textquotedblright\ \textit{Computing in
Science and Engineering}, 2, 22-23.

\item Hastings, W. K. (1970) \textquotedblleft Monte Carlo Sampling Methods
using Markov Chains and their Applications,\textquotedblright%
\ \textit{Biometrika,} 57, 97--109.

\item Liu, J. S. (2001) \textit{Monte Carlo Strategies in Scientific
Computing, }New York: Springer-Verlag.

\item Metropolis, N.; Rosenbluth, A. W.; Rosenbluth, M. N.; Teller A. and
Teller H. (1953) \textquotedblleft Equation of State Calculations by Fast
Computing Machines,\textquotedblright\ \textit{The Journal of Chemical
Physics,} 21, 1087-1092.

\item Minh, D. L. (2001) \textit{Applied Probability Models}, Pacific Grove,
CA: Duxbury Press.

\item Minh, D. L.; Minh, D. D. L. and Nguyen A. (2012) \textquotedblleft
Regenerative Markov Chain Monte Carlo for Any Distribution,\textquotedblright%
\ \textit{Communications in Statistics - Simulation and Computation}, 41, 1745-1760.

\item Peskun, P. H. (1973). \textquotedblleft Optimum Monte Carlo Sampling
using Markov chains,\textquotedblright\ \textit{Biometrika,} 60, 607--612.

\item Tierney, L. (1994) \textquotedblleft Markov Chains for Exploring
Posterior Distributions,\textquotedblright\ \textit{The Annals of Statistics},
22, 1701--1728.

\item von Neumann (1951) \textquotedblleft Various Techniques Used in
Connection with Random Digits,\textquotedblright\ \textit{National Bureau of
Standards}, Applied Mathematics Series, 12, 36-38.
\end{enumerate}

\end{document}